\documentstyle[aps,epsf,prl,multicol]{revtex}

\begin{document}
\title{
Can disorder induce a finite thermal conductivity in 1D lattices?
}
\author{
Baowen Li$^{1,2}$,
Hong Zhao,$^{2,3}$ and Bambi Hu$^{2,4}$
} 
\address{
$^{1}$ Department of Physics, National University of Singapore, 
119260 Singapore \\
$^{2}$Department of Physics and Center for Nonlinear Studies, Hong 
Kong Baptist University, China\\
$^{3}$ Department of Physics, Lanzhou University, 730000 Lanzhou, China\\
$^{4}$ Department of Physics, University of Houston, Houston TX
77204-506\\
}
\date{\today} 
\maketitle

\begin{abstract} 
We study heat conduction in one dimensional mass
disordered harmonic and anharmonic lattices.
It is found that the thermal conductivity $\kappa$ of the disordered anharmonic lattice is 
finite at 
low temperature,  whereas it diverges as $\kappa \sim N^{0.43}$ at high temperature.
Moreover, we demonstrate that a unique nonequilibrium stationary state  in the  
disordered harmonic lattice does not exist at all. 
\end{abstract} 

\begin{multicols}{2}

Can disorder induce a finite thermal conductivity in one dimensional
(1D) lattices? This question arose immediately after Anderson's 
finding of localization\cite{Anderson58}. It has been commonly believed that
disorder scatter normal modes and induce a diffusive 
energy transport that leads
to a normal heat conduction. However, the early
numerical as well as theoretical studies shown that the thermal
conductivity in 1D (mass) disordered harmonic lattice is proportional to
$N^{1/2}$, where $N$ is the length of the
lattice\cite{Payton,Disorder,Ishii}. The reason for this divergent thermal
conductivity is not clear up to now. In fact, the existing 
theories\cite{Disorder,Ishii} do not prove that
a unique nonequilibrium stationary state
could be reached in a disordered harmonic lattice.
Intended to obtain a finite thermal
conductivity, Payton {\it et al} \cite{Payton} tried to add an anharmonicity
to the disordered lattice. It was found that the anharmonicity did enhance
the heat current, but Payton {\it et al.} were not able to verify the 
validity of the Fourier law in such systems due to limited computer 
facility. Therefore, whether disorder can induce a finite thermal conductivity
is still a puzzle. 

In past three decades, many works have been devoted to seek 
such a model whose heat conduction behavior obeys the Fourier law, 
i.e. $J= -\kappa dT/dx$. The 
primary motivation is to find out whether it is possible to prove and/or 
disprove a given 1D many-body Hamiltonian system having or not having a 
finite thermal conductivity. If it is possible, then how can one calculate 
transport coefficients from microscopic Hamiltonian.
The emergence of chaos in past years sheds some lights on this old 
but rather fundamental problem. It stimulated further 
study of heat conduction in non-integrable systems, aims at  
finding dynamical origin of the thermodynamical properties, such as 
irreversibility etc.\cite{chaos98}. For this purpose, a wide range of 
non-integrable systems such as the Lorenz gas model \cite{Lorenz}, 
the ding-a-ling and alike models\cite{Casati84}, the 
Fermi-Pasta-Ulam (FPU) model\cite{FPU}, the Frenkel-Kontoroval model\cite{HLZ98}, 
the diatomic Toda lattice\cite{Hatano99}, the 
Heisenberg spin lattice\cite{Dhar99}, the $\phi^4$ lattice\cite{HLZ99} and the sinh-Gordon and bounded single-well models\cite{Tsironis} etc, have been invoked. Most recently, the study has been extended to systems
having periodic inter-particle potential
\cite{period00}.
However, as we have already pointed out \cite{HLZ99} that 
non-integrability is only a
{\it necessary} condition for the formation of a temperature gradient but 
{\it not a sufficient} one for a finite thermal conductivity.

A rigorous proof of the necessary and sufficient condition for the Fourier 
law is still lacking, even though some progress in this direction has been achieved\cite{Prosen00}. 
Facing up such a situation, we have to rely on large 
scale molecular dynamics simulations to obtain a deep insight of  
underlying mechanism of the divergent and/or finite thermal conductivity.
Recently\cite{HLZ99}, in searching for the underlying mechanism of the energy transport,
we found that the (mass) uniform  
lattice systems can be classified into three categories: integrable one, non-integrable 
one with on-site potential, and non-integrable one without on-site potential. 
We discovered that the on-site potential plays a very important role. It destroys 
the momentum conservation and produces a diffusive energy transport, and
thus leads to a finite thermal conductivity. In non-integrable lattices 
without on-site potential, the scattering of solitary waves is found to be 
responsible for the divergent thermal conductivity $\kappa\sim N^{0.43}$ in
the FPU and alike models.

In this Letter, we would like to study the effect of disorder on heat conduction in 1D lattices.
The Hamiltonian of the models we are considering can be written as
\begin{equation}
H = \sum_i H_i,\qquad H_i=\frac{p_{i}^{2}}{2m_i} + V(x_{i-1},x_i),
\label{Ham}
\end{equation}
where $V(x_{i-1},x_i)$ stands for the interaction potential of the 
nearest-neighbor particles. In our study the mass of particles is 
given by: 
\begin{equation}
m_i = m_0 + \lambda (R_i - 0.5),
\label{Randmass}
\end{equation}
where $\lambda$ is a parameter adjusting the amplitude of the random mass, and
$R_i$ is a random number distributed uniformly in the interval of $(0,1)$. 

{\it  Disordered  anharmonic lattices}.  In this case the inter-particle potential 
takes the form as that in the FPU model, 
\begin{equation}
V(x_{i-1}, x_i) = \frac{1}{2}(x_i - x_{i-1})^2  + \frac{\beta}{4}\left(x_i - 
x_{i-1}\right)^4.
\label{FPUV}
\end{equation}
$\beta=1$ throughout this paper except otherwise stated. We call the model  
disordered FPU (DFPU) model.
In our numerical simulations the Nos\'e-Hoover thermostats \cite{NH85}
were put on the first and the last particles, keeping them at temperature
$T_+$ and $T_-$, respectively.  The eighth-order Runge-Kutta algorithm was
used to solve the coupled differential equations. All computations were
carried out in double precision.  Usually the stationary state set in
after $10^6 \sim 10^7$ time units, thereafter the time average $\langle 
J_i\rangle$
($J_i =\dot{x}_i \partial V/\partial x_{i+1}$) was found to be site
independent, and thus is denoted by $J$. 

\begin{figure} 
\epsfxsize=8cm 
\epsfbox{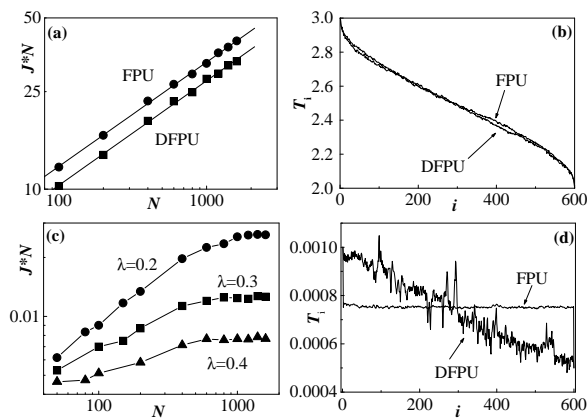} 
\vspace{0cm}
\narrowtext 
\caption{$JN$ versus $N$ (a) and temperature profiles (b) for DFPU
 model ($\lambda =0.4$) and the
FPU model at high temperature, $T_+=3$ and $T_-=2$. The solid 
lines in (a) are the best fitting ones, both have same slope 
$\alpha = 0.43\pm 0.01$. (c) $JN$ versus $N$
for the DFPU model with different values of disorder at low temperature,
$T_+=10^{-3}$ and $T_-=5\times10^{-4}$. The lines are drawn to guide 
eye. (d) Same as (b) but for the case of low temperature.} 
\label{JN}
\end{figure}

In Fig. \ref{JN}(a), we present $JN$ versus $N$ for the FPU
and the DFPU models with $T_+=3$ and $T_- = 2$. 
Both models have a similar (almost 
indistinguishable) temperature profile (Fig. \ref{JN}(b)).
In both cases, the
temperature gradient is proportional to $N^{-1}$, thus $JN 
\propto \kappa$. Best fitting in Fig.\ref{JN}(a) 
yields $\kappa \sim N^{0.43\pm0.01}$. This implies that in both models the thermal 
conductivity diverges in the same way. More numerics show that in this high temperature regime, the disorder 
does not help converge the thermal conductivity, it only decreases the value of 
heat current.

Interesting things turn out when
the temperature of the thermostats is decreased.  In Fig.\ref{JN}(c), $JN$ versus $N$ is
plotted for the DFPU model at three different values of disorder
$\lambda=0.2, \,0.3$, and $0.4$, respectively, with
$T_+=10^{-3}$ and $T_-=5\times 10^{-4}$.
When $N< N_c$,
$JN$ increases monotonically with $N$. 
When $N > N_c$, $JN$ saturates, namely $\kappa$ becomes an $N$-independent
constant. This result indicates that the heat conduction obeys Fourier law 
when the size of the 1D lattice becomes larger than $N_c$.
However, in this temperature regime,
the FPU model and the harmonic lattice have 
the same heat conduction behavior \cite{HLZ99}. This is reflected in the temperature 
profile shown in Fig. \ref{JN}(d). It demonstrates that
no temperature gradient 
can be formed, and the stationary state corresponds to $T=(T_++T_-)/2$.

\begin{figure}
\epsfxsize=8cm
\epsfbox{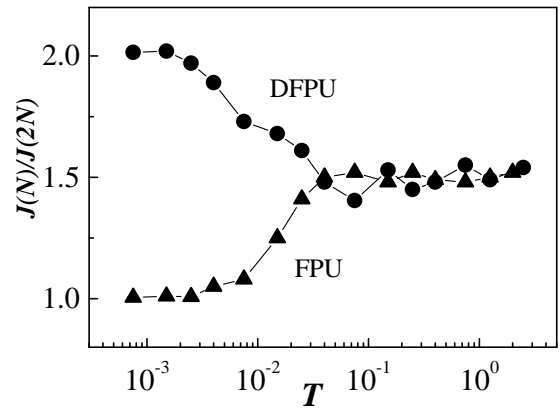}
\vspace{-4cm}
\narrowtext
\caption{$J(N)/J(2N)$ (for $N=600$) versus the 
average 
temperature $T = (T_++T_-)/2$ in the semi-logarithmic scale for the DFPU (solid circle)
($\lambda =0.4$) and the FPU (solid triangle) models. The FPU model shows a transition to a
harmonic lattice with the decrease of the temperature. However,
the DFPU model changes from a 
divergent thermal conductor to a normal one as the temperature is decreased.
} 
\label{Flux}
\end{figure} 

In Fig.\ref{Flux}, we show $J(N)/J(2N)$ versus 
$T=(T_++T_-)/2$ for the FPU (solid triangle) and the DPFU (solid circle) models. Where $J(N)$ 
is the heat current for the lattice of length $N$. 
The transition of the FPU model to a harmonic lattice happens
when the temperature is decreased below a threshold value of $T_c$ ( $\approx 10^{-2}$). 
This is illustrated in Fig. 2 by an abrupt decrease of $J(N)/J(2N)$ from 1.5 to 1. 
On the other hand, within almost the same temperature range the heat conduction in the DFPU model undergoes another transition. This transition is demonstrated by an abrupt 
increase of $J(N)/J(2N)$ from about 1.5 to 2 when $T<T_c$. This implies that the DFPU model transforms from an abnormal thermal conductor
to a normal one.

The above results of the FPU model can be understood from the scattering 
mechanism of solitary waves. As we discovered \cite{HLZ99} that the scattering
between solitary waves from opposite directions causes an energy loss
that gives rise to a temperature gradient.  However, the momentum
conservation in the FPU model prohibits a diffusive energy transport.  The interaction of the solitary
waves depends on the initial excitation. When
$T < T_c$, the solitary waves cannot be excited any more,
the energy is transported by linear excitations - phonons, thus the
FPU model behaves like a harmonic one. 

\begin{figure}
\epsfxsize=8cm
\epsfbox{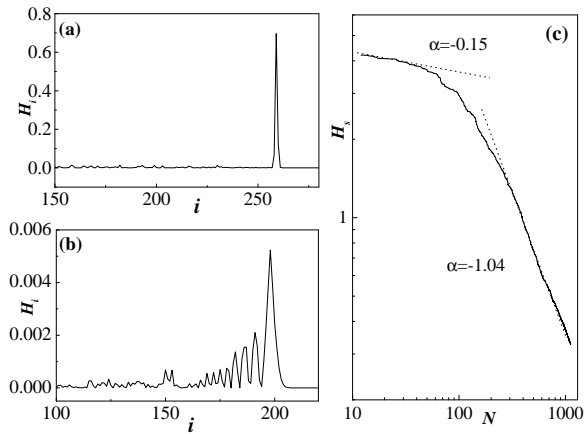}
\vspace{0.cm}
\narrowtext
\caption{
Energy distribution in space at  time $t=200$ for high 
excitation - solitary waves (a) and low excitation - phonons (b) for a 
DFPU lattice with
disorder $\lambda=0.4$.  (c) The energy of solitary 
waves on different site. The dotted lines are the best fitting ones with 
the data in initial stage and in final stage, respectively. The different slopes 
(given in the figure) show that the energy decays in different rate for 
solitary waves (initial stage) and phonons (final stage).
} 
\label{Energy} 
\end{figure}

In order to clarify the underlying mechanism of the heat conduction in the
disordered model, we study the dynamics of a single excitation.
Figure \ref{Energy}(a)  shows a snapshot of energy density $H_i$
in a DFPU lattice at $t=200$ for an initial excitation on the left
end, i.e. $p_1=3, p_i=0$, for $i\ne 1 $. A solitary wave is excited and
propagates to $i=250$. If we decrease the energy of initial excitation
below a threshold value, we find that the energy is no longer localized,
namely the solitary wave disappears. Instead, a traveling wave packet, 
linear excitations, is created and moves diffusively (see Fig.
\ref{Energy}(b)).  However, unlike in the case of the FPU model, when a
single solitary wave or a linear wave packet travels along the disordered
lattice, its energy will decrease due to the scattering from the disorder.
An interesting phenomenon is that solitary waves and phonons obey different scattering laws,
as is shown in Fig. \ref{Energy}(c). In this
figure we plot the energy versus the distance of the solitary wave
excited in Fig.\ref{Energy}(a). We see that both in initial
and final stages, the energy decays in power law $H_s\sim N^{-\alpha}$ but with different exponents, $\alpha \approx 0.15$ in the initial stage and $\alpha \approx 1$ in the final stage. The former represents the scattering of solitary waves by disorder, and the latter represents the scattering of 
phonons by disorder as in the final stage the energy becomes so low that the solitary waves cannot be excited any more.

With above picture, we can explain the numerical results of
the DFPU model shown in Figs. \ref{JN}-\ref{Flux}. At high 
temperature (Fig. \ref{JN}(a-b) and Fig. \ref{Flux}), 
energy is transported mainly by solitary waves which 
encounter two kinds of scattering: scattering from solitary waves and scattering
from disorder. The scaling law of energy loss caused by the two
scattering are $N^{-\alpha}$ but with different values of $\alpha$,
$\alpha \approx 0.5$ for the former\cite{HLZ99}, $\alpha \approx 
0.15$ for the latter (see Fig 3(c)). In the thermodynamic limit the 
main cause of energy 
loss comes from the scattering of solitary waves. 
Thus at high temperature the DFPU model shows the same scaling law as
the FPU model as is shown in Fig \ref{JN}(a). In this case the disorder 
cannot induce a finite thermal conductivity, it decreases the total heat current only.

The situation becomes different at low temperature (Fig. \ref{JN} (c-d) and  
Fig.\ref{Flux}).
In this case, the
solitary wave is hardly excited. The energy is transported by phonons. The 
scattering of phonons by disorder becomes a
dominant factor for energy loss. Thus the energy decays as $N^{-1}$ as is shown in Fig.
\ref{Energy}(c), which implies a finite thermal conductivity.
This prediction
is confirmed by a direct calculation of $JN$ given in Fig. \ref{JN}(c).  
It is natural to relate the saturation phenomenon of $JN$ in Fig. \ref{JN}(c) to the localization of phonons in the system,
wherein the localization length $l$ is an important scale. If the length of
the lattice $N$ is shorter than $l$ the thermal conductivity is 
expected to be
divergent. On the contrary, if the lattice is much longer than $l$ then a
normal heat conduction could be realized. This is indeed the case 
shown in Fig.\ref{JN}(c).  The threshold value $N_c$ is approximately
$500, 700$, and $1000$ for the three different values of the disorder
$\lambda=0.4, 0.3,$ and $0.2$, respectively. Extensive numerical simulations
confirm that $N_c \propto 1/\lambda$. This is of order $l$,
because $l \propto \langle m\rangle/\sqrt{\langle (m-\langle
m\rangle)^2\rangle} \sim 1/\lambda$\cite{Ishii}. 

{\it Disordered harmonic lattices}. 
It was shown theoretically \cite{Disorder,Ishii} that the thermal conductivity diverges 
as $\kappa \sim N^{1/2}$  in 1D disordered harmonic lattice 
($\beta=0$ in Eq.(\ref{FPUV})). The proof was based on an assumption that a 
unique nonequilibrium stationary state could be reached in such systems.
However, to the best of our knowledge, this assumption has not been verified numerically.
This motivated us to do further investigation.

\begin{figure}
\epsfxsize=8cm
\epsfbox{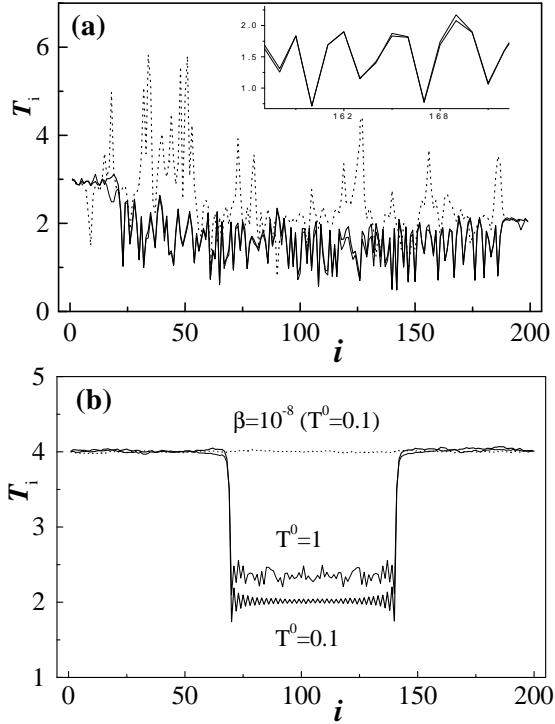}
\vspace{0cm}
\narrowtext
\caption{(a) Temperature distribution of a disordered 
harmonic 
lattice ($\lambda=1$) from  different initial conditions, the solid lines for $T^0=0.01$ 
at $10^6$ and $10^7$ time units (see also the inset for the 
magnification of a small interval), respectively, and the dotted line
for $T^0=3$ at $10^7$ time units. (b) The temperature 
profile (solid line) of a harmonic 
lattice consists of three uniform segments. The left and the right 
segments are atoms of the mass $m=1$ while the middle one consists of atoms of 
mass $m=0.5$. $T_+ = T_-=4$.  The dashed line is for the same system but with a tiny anharmonicity $\beta = 10^{-8}$.
} 
\label{Harmd} 
\end{figure}

We find that a stationary state, with erratic temperature fluctuation, can be
set up as is shown in Fig.  \ref{Harmd}(a). The states starting from
the same initial condition $(T^0=0.01$) are almost indistinguishable at
different time $t=10^6$ and $t=10^7$. (Please note the inset for magnification.) 
This guarantees the existence of the stationary state. Unfortunately, such a
state depends sensitively on initial condition . In the same figure, we
show another stationary state (dashed line) formed from a different initial condition 
$T^0=3$. It differs greatly from that one of $T^0=0.01$. This result means that the nonequilibrium stationary state in the disordered harmonic lattice is {\it not unique}.

To obtain a clear picture, we turn to a simple model, a lattice
consists of only two types of atoms, one with mass $m=1$ the other with
mass $m=0.5$. A segment of light atoms is embedded in
the middle of other two segments of heavy atoms. Two
ends of the lattice are put in the thermostats with the same
temperature $T_+=T_-=4$. Two stationary states from two different
initial conditions $T^0=0.1$ and $T^0=1$ are shown in Fig.\ref{Harmd}(b). 
Again, no unique stationary state can be formed. However,
adding a small fraction of anharmonicity, $\beta=10^{-8}$, to this lattice produces a 
complete different picture. With this tiny anharmonicity, we can obtain not only 
 a smooth temperature profile but also a unique stationary 
state, see Fig. \ref{Harmd}(b).

The above results show that although the anharmonicity 
alone is not enough to yield a finite thermal conductivity, it plays a
crucial role in establishing a unique 
nonequilibrium stationary state in disordered lattices.

In summary, we have studied heat conduction in 1D disordered
harmonic and anharmonic lattices.  Our numerical results
show that at low temperature disorder can induce a finite thermal conductivity.
The magnitude of
the disorder does not affect the results in the thermodynamical limit. It
only determines the localization length and threshold length of a lattice
having a finite thermal conductivity. However, at high temperature, the 
disordered anharmonic lattice shows a divergent thermal conductivity, 
which is similar to that of 
the FPU model. In addition, we provide a numerical evidence 
showing that the nonequilibrium stationary state in a disordered harmonic lattice 
is {\it not unique}.

\bigskip
BL was supported in part by Academic Research Fund of 
National University of Singapore.
This work was also supported in part by Hong 
Kong Research Grant Council and the Hong Kong Baptist University Faculty
Research Grant.


\end{multicols}
\end{document}